# ASTROPHYSICAL RADIATIVE NEUTRON CAPTURE ON $^{10}$B TAKING INTO ACCOUNT RESONANCE AT 475 keV

Sergey Dubovichenko[1,2,†] and Albert Dzhazairov-Kakhramanov[1,2,‡]

[1]*V. G. Fessenkov Astrophysical Institute "NCSRT" NSA RK, 050020, Observatory 23, Kamenskoe plato, Almaty, Kazakhstan*
[2]*Institute of Nuclear Physics CAE MINT RK, 050032, str. Ibragimova 1, Almaty, Kazakhstan*
[†]*dubovichenko@mail.ru*, [‡]*albert-j@yandex.ru*

## ABSTRACT

The possibility of the description of the available experimental data for cross sections of the neutron capture reaction on $^{10}$B at thermal and astrophysical energies, taking into account the resonance at 475 keV, was considered within the framework of the modified potential cluster model with forbidden states and accounting for the resonance behavior of the scattering phase shifts.

*Subject Keywords*: Nuclear astrophysics; primordial nucleosynthesis; light atomic nuclei; low and astrophysical energies; radiative capture; thermonuclear processes.

## 1. Introduction

One extremely successful line of development of nuclear physics in the last 50 years has been the microscopic model known as the Resonating Group Method (RGM, see, e.g., Wildermut & Tang 1977; Mertelmeir & Hofmann 1986), and the associated Generator Coordinate Method (see, particularly, Descouvemont & Dufour 2012) or algebraic version of RGM (see, e.g., Nesterov et al. 2010). Such evident success has led the majority of physicists to the view that the advancement of new results in low-energy nuclear physics and nuclear astrophysics is possible only in this direction. Eventually, a very common but evidently wrong opinion states that this is the only way in which future development of our ideas about the structure of the atomic nucleus, and nuclear and thermonuclear reactions at low and astrophysical energies can be imagined.

However, the possibilities offered by a simple two-body potential cluster model (PCM) have not been studied fully up to now, particularly if it uses the concept of forbidden states (FS), see, e.g., Nemets et al. (1988), and considers directly the resonance behavior of the elastic scattering phase shifts of interacting particles at low energies (see, e.g., Dubovichenko 2012a, 2012b). Such a model can be called a modified PCM (MPCM). The rather difficult RGM calculations are not the only way in which to explain the available experimental facts. The simpler MPCM with FSs can be used by taking into account the classification of orbital states according to the Young tableaux and the resonance behavior of the elastic scattering phase shifts. In many cases, such an approach, as has been shown previously (see, e.g., Dubovichenko 2012a, 2012b), allows one to obtain adequate results in the description of many experimental data for the total cross sections of the thermonuclear capture process.

Therefore, in continuing to study the processes of radiative capture (see, e.g., Dubovichenko 2012a, 2012b), we will consider the n+$^{10}$B → $^{11}$B+γ reaction within the framework of the MPCM at low and thermal energies. The resonance behavior of the elastic scattering phase shifts of the interacting particles at low energies will be taken into account. In addition, the classification of the orbital states of the clusters according to the Young tableaux allows one to clarify the number of FSs and allowed states (ASs), i.e., the number of nodes of the wave function of the relative motion of the cluster. The potentials of the n$^{10}$B interaction for scattering processes will be constructed based on the reproduction of the spectra of resonance states for the final nucleus in the n$^{10}$B channel. The n$^{10}$B potentials are constructed based on the description both of the binding energies of these particles in the final nucleus and of certain basic characteristics of these states; for example, the charge radius and the asymptotic constant (AC) for the bound state (BS) or the ground state (GS) of $^{11}$B, formed as a result of the capture reaction in the cluster channel, which coincide with the initial particles (see, e.g., Dubovichenko 2012b).

The study of the reaction $^{10}$B(n, γ)$^{11}$B, from the astrophysical point of view, is interesting because the resultant $^{11}$B is a part of the reaction chains in the so-called inhomogeneous Big Bang models (see, e.g., Heil et al. 1998; Guimaraes & Bertulani 2010; Igashira & Ohsaki 2004; Nagai et al. 1996; Liu et al. 2001). Therefore, it is interesting to consider the additional chain of these reactions starting from the boron isotope $^{10}$B:

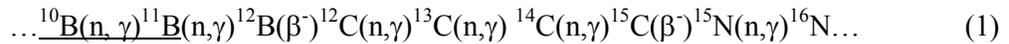

$$\ldots {}^{10}\text{B(n, γ)}{}^{11}\text{B(n,γ)}{}^{12}\text{B(β}^-){}^{12}\text{C(n,γ)}{}^{13}\text{C(n,γ)}\ {}^{14}\text{C(n,γ)}{}^{15}\text{C(β}^-){}^{15}\text{N(n,γ)}{}^{16}\text{N}\ldots \qquad (1)$$

Evidently, the considered reaction can play a certain role in some models of the Universe, when the number of forming nuclei, perhaps, is dependent on the presence of dark energy and its concentration (see, e.g., Esmakhanova et al. 2011), on the rate of growth of baryonic matter perturbations (see, e.g., Chechin 2006), or on the rotation of the early Universe (see, e.g., Chechin 2010). However, perturbations in the primordial plasma not only stimulate the process of nucleosynthesis (see, e.g., White Scott & Silk 1994), but also kill it, for example, through the growth of the perturbations of nonbaryonic matter of the Universe (see, e.g., Chechin & Myrzakul 2009) or because of the oscillations of cosmic strings (see, e.g., Omarov & Chechin 1999).

In addition, to our mind, this study has great importance because it has been impossible for us to find any similar theoretical calculations for the reaction $^{10}$B(n, γ)$^{11}$B in the thermal and astrophysical energy range.

## 2. Structure of cluster states

We regard the results of the classification of $^{11}$B by orbital symmetry in the n$^{10}$B channel as qualitative, because there are no complete tables of Young tableaux productions for systems with more than eight nucleons (see, e.g., Itzykson & Nauenberg 1966), which have been used in earlier similar calculations (see, e.g., Dubovichenko 2012a, 2012b). At the same time, simply based on such a classification, we succeeded in describing the available experimental data on the radiative capture of neutrons and charged particles for a wide range of reactions (see, e.g., Dubovichenko 2012a, 2012b; Dubovichenko, Dzhazairov-Kakhramanov & Burkova 2013; Dubovichenko & Dzhazairov-Kakhramanov 2012a; Dubovichenko & Uzikov 2011). This is why the classification procedure by orbital symmetry given above was used here for the determination of the number of FSs and ASs in partial intercluster potentials and, consequently, to the specified number of nodes of the wave function of the relative motion of the cluster for the case of neutrons and $^{10}$B.

Furthermore, we will suppose that it is possible to assume the orbital Young tableau in the form {442} for $^{10}$B; therefore, for the n$^{10}$B system, we have {1} × {442} → {542} + {443} + {4421} (see, e.g., Itzykson & Nauenberg 1966). The first of the obtained tableaux is compatible with orbital moments $L = 0, 2, 3$, and $4$, and is forbidden because it contains five nucleons in the $s$-shell. The second tableau is allowed and is compatible with orbital moments $L = 1, 2, 3$, and $4$, and the third is also allowed and is compatible with $L = 1, 2$, and $3$ (see, e.g., Neudatchin & Smirnov 1969). As mentioned before, the absence of tables of Young tableaux productions for when the number of particles is 10 and 11 prevents the exact classification of the cluster states in the considered system of particles. However, qualitative estimations of the possible Young tableaux for orbital states allow us to detect the existence of the FSs in the *S* and *D* waves and the absence of FSs for the *P* states. The same structure of FSs and ASs in the different partial waves allows us to construct the potentials of intercluster interactions required for the calculations of the total cross sections for the considered radiative capture reaction. Thus, by limiting our consideration to only the lowest partial waves with orbital moment $L = 0$, and 1, it could be said that for the n$^{10}$B system (for $^{10}$B it is known $J^{\pi}, T = 3^{+}, 0$; see, e.g., Tilley et al. 2004), the only allowed state exists in the *P* wave potentials and the FS is in the *S* waves. The state in the $^{6}P_{3/2}$ wave (representation in $^{(2S+1)}L_J$) corresponds to the GS of $^{11}$B with $J^{\pi}, T = 3/2^{-}, 1/2$ and is at the binding energy of the n$^{10}$B system of -11.4541(2) MeV (see, e.g., Ajzenberg-Selove 1990). Let us note that some n$^{10}$B scattering states and BSs can be mixed by isospin with $S = 5/2$ ($2S+1 = 6$) and $S = 7/2$ ($2S + 1 = 8$).

The spectrum of $^{11}$B for excited states (ESs), bound in the n$^{10}$B channel, shows that at the energy of 2.1247 MeV above the GS or -9.3329 MeV (Ajzenberg-Selove 1990) relative to the threshold of the n$^{10}$B channel, the first ES can be found, bound in this channel with the moment $J^{\pi} = 1/2^{-}$, which can be compared with the $^{6}F_{1/2}$ wave with an FS. However, we will not consider it, because of the large value of the angular momentum barrier. The second ES at the energy 4.4449 MeV (Ajzenberg-Selove 1990) above the GS or -7.0092 MeV relative to the threshold of the n$^{10}$B channel has the moment $J^{\pi} = 5/2^{-}$, and it can be compared with the mixture of the $^{6}P_{5/2}$ and $^{8}P_{5/2}$ waves without FSs. Furthermore, the unified potential of such a mixed $P_{5/2}$ state will be constructed, because the model used does not allow one to divide states with different spin clearly. The wave function obtained with this potential at the calculation of the Schrödinger equation and, in principle, consisting of two components for different spin channels, does not divide into these components in the explicit form, i.e., we are using the total form of the WF in all calculations. The third ES at the energy of 5.0203 MeV (Ajzenberg-Selove 1990) relative to the GS or -6.4338 MeV relative to the channel threshold has the moment $J^{\pi} = 3/2^{-}$, and it can be matched to the $^{6}P_{3/2}$ wave without an FS. The fourth ES at the energy of 6.7429 MeV (Ajzenberg-Selove 1990) relative to the GS or -4.7112 MeV relative to the channel threshold has the moment $J^{\pi} = 7/2^{-}$, and it can be matched to the mixture of the $^{6}P_{7/2}$ and $^{8}P_{7/2}$ waves without an FS. In addition, it is possible to consider the ninth ES at the energy of 8.9202 MeV with the moment 5/2$^{-}$, i.e., at the energy of -2.5339 MeV relative to the n$^{10}$B threshold, which can be matched to the mixture of the $^{6}P_{5/2}$ and $^{8}P_{5/2}$ states without FSs.

Consider now the resonance states in the n$^{10}$B system, i.e., states at positive energies. The first resonance state of $^{11}$B in the n$^{10}$B channel, located at the energy 0.17 MeV, has the neutron width of 4 keV and the moment $J^{\pi} = 5/2^{+}$ (see, e.g., Ajzenberg-Selove 1990). It is possible to compare this state with the $^{6}S_{5/2}$ scattering wave with an FS. We have not succeeded in the construction of the potential with such small width; therefore, we will consider this scattering wave as non-resonant, which leads to zero scattering phase shifts. The second resonance state has the energy of 0.37 MeV – its neutron width equals 0.77 MeV and the moment $J^{\pi} = 7/2^{+}$ (Ajzenberg-Selove 1990); therefore, it is possible to compare this with the $^{8}S_{7/2}$ scattering wave with an FS. Because of the large width of resonance (two times greater than its energy), we will use non-resonant values of the parameters for the potential to coincide with the previous $^{6}S_{5/2}$ potential. The third resonance state has the energy of 0.53 MeV in the laboratory system (l.s.) – its neutron width equals 0.031 MeV in the center-of-mass system (c.m.) and the moment $J^{\pi} = 5/2^{-}$ (Ajzenberg-Selove 1990). Therefore, it can be compared with the mixed $^{6}P_{5/2} + ^{8}P_{5/2}$ scattering waves without FSs. These characteristics of the resonance are given in Table 11.11 of work Ajzenberg-Selove (1990), and in the note to this table are given the energy and the width values equal to 0.495(5) MeV and 140(15) keV, respectively, with reference to Lamaze, Schrack & Wasson (1978). At the same time, the value of 0.475(17) MeV (l.s.) with the total width 200(20) keV (c.m.) is given in Table 11.3 of Ajzenberg-Selove (1990) for this resonance. Furthermore, under the construction of this potential, we will proceed from two variants of data, notably, the first and the last given above with the width of 31 keV (c.m.), but at the energy of the resonance of 475 keV (l.s.), which follows from the level spectrum (Ajzenberg-Selove 1990).

The next resonance is at the energy above 1 MeV and we will not consider it (see Table 11.11, Ajzenberg-Selove 1990). There are no resonance levels lower than 1 MeV in the spectrum of $^{11}$B that can be matched to the $^{6}P_{3/2}$ and $^{6+8}P_{7/2}$ states (see, e.g., Ajzenberg-Selove 1990). Therefore, their phase shifts are taken as equal to zero, and as far as there are no FSs in the $P$ waves, by way of the first variant, such potentials can be simply equalized to zero (see, e.g., Dubovichenko 2012a, 2012b). We ought to note here that there are more up-to-date values for all these states (see, e.g., Kelley et al. 2012) – the results from this review do not differ for the ESs (see Table 11.18, Kelley et al. 2012), but do have slightly different values for resonance states. Particularly, the excited energy of 11.893(13) MeV with the adjusted total width of 194(6) keV, which gives 483 keV (l.s.) for the resonance energy, is given for the state $J^\pi = 5/2^-$, which can be matched to the $^{6+8}P_{5/2}$ scattering waves without FSs. The width equal to 1.34 MeV, which is given for the state with $J^\pi = 7/2^+$, compares with the $^{8}S_{7/2}$ scattering wave with an FS – it is twice that given in Ajzenberg-Selove (1990).

Continuing to the analysis of possible electromagnetic $E$1 and $M$1 transitions, let us note that we will consider only transitions to the GS and to four (2$^{nd}$, 3$^{rd}$, 4$^{th}$, and 9$^{th}$) ESs from the $S$ and $P$ scattering waves. As the GS is matched with the $^{6}P_{3/2}$ level, it is possible to consider $E$1 transitions from the $^{6}S_{5/2}$ scattering wave to the GS of $^{11}$B.

No. 1. $^{6}S_{5/2} \to ^{6}P_{3/2}$.

In addition, it is possible to consider $E$1 transitions from the $^{6}S_{5/2}$ and $^{8}S_{7/2}$ scattering waves to the second ES of $^{11}$B, which is the mixture of two $P$ states:

No. 2. $\begin{array}{l}^{6}S_{5/2} \to ^{6}P_{5/2} \\ ^{8}S_{7/2} \to ^{8}P_{5/2}\end{array}$.

Because here we have transitions from the initial $S$ states that differ by spin to the different parts of one total WF of the BS, which, evidently, have no essential difference, the cross section of these transitions will sum up, i.e., $\sigma = \sigma(^{6}S_{5/2} \to ^{6}P_{5/2}) + \sigma(^{8}S_{7/2} \to ^{8}P_{5/2})$. The third ES is matched with the $^{6}P_{3/2}$ level as the GS, and it is possible to consider the $E$1 transitions from the $^{5}S_{5/2}$ scattering waves to this ES of $^{11}$B.

No. 3. $^{6}S_{5/2} \to ^{6}P_{3/2}$.

Another $E$1 transition is possible from the $^{6}S_{5/2}$ and $^{8}S_{7/2}$ scattering waves of the fourth ES of $^{11}$B at $J^\pi = 7/2^-$:

No. 4. $\begin{array}{l}^{6}S_{5/2} \to ^{6}P_{7/2} \\ ^{8}S_{7/2} \to ^{8}P_{7/2}\end{array}$.

The cross section of these two transitions will also sum up. The last of considered $E$1 transitions is the capture from the $^{6}S_{5/2}$ and $^{8}S_{7/2}$ scattering waves to the ninth ES of $^{11}$B at $J^\pi = 5/2^-$:

No. 5. $\begin{array}{l}^{6}S_{5/2} \to ^{6}P_{5/2} \\ ^{8}S_{7/2} \to ^{8}P_{5/2}\end{array}$.

The cross section of these transitions will also sum up, as given in the case of reaction 2.

Furthermore, it is possible to consider *M*1 transitions to the GS from the resonance scattering wave $^6P_{5/2}$ at 0.475(17) MeV, and from the non-resonance $^6P_{3/2}$ wave.

No. 6. $\begin{aligned}^6P_{5/2} &\to {}^6P_{3/2}\\ {}^6P_{3/2} &\to {}^6P_{3/2}\end{aligned}$.

As will be shown later, the cross section of the transition $^6P_{3/2} \to {}^6P_{3/2}$ for the first variant of the $^6P_{3/2}$ scattering potential with zero depth will stay at the level 1–2 µb, and the other transitions from the non-resonance waves will not be considered. Furthermore, it is possible to consider *M*1 transitions to the second ES $^6P_{5/2}$ and $^8P_{5/2}$ from the resonance $^6P_{5/2}$ and $^8P_{5/2}$ scattering waves.

No. 7. $\begin{aligned}^6P_{5/2} &\to {}^6P_{5/2}\\ {}^8P_{5/2} &\to {}^8P_{5/2}\end{aligned}$.

Because this is the transition from the mixed-by-spin $P_{5/2}$ scattering wave to the mixed-by-spin second ES, the cross section will be averaged according to transitions given above, i.e., σ = 1/2 {σ($^6P_{5/2} \to {}^6P_{5/2}$) + σ($^8P_{5/2} \to {}^8P_{5/2}$)}. The possible transition $^{6+8}P_{7/2} \to {}^{6+8}P_{5/2}$ from the non-resonance $^{6+8}P_{5/2}$ wave to the second ES is not taken into account here.

It is possible to consider the *M*1 transitions to the third ES $^6P_{3/2}$ from the resonance $^6P_{5/2}$ scattering wave

No. 8. $^6P_{5/2} \to {}^6P_{3/2}$.

The *M*1 transitions are feasible to the fourth ES $^6P_{7/2} + {}^8P_{7/2}$ from the resonance $^6P_{5/2} + {}^8P_{5/2}$ scattering wave

No. 9. $\begin{aligned}^6P_{5/2} &\to {}^6P_{7/2}\\ {}^8P_{5/2} &\to {}^8P_{7/2}\end{aligned}$.

This cross section will be averaged over two transitions, as was given in the case of reaction 7.

Finally, we can consider the *M*1 transitions to the ninth ES $^6P_{5/2} + {}^8P_{5/2}$ from the resonance $^6P_{5/2} + {}^8P_{5/2}$ scattering wave, because there are experimental data (see, e.g., Igashira et al. 1994) for this.

No. 10. $\begin{aligned}^6P_{5/2} &\to {}^6P_{5/2}\\ {}^8P_{5/2} &\to {}^8P_{5/2}\end{aligned}$.

This cross section will also be averaged over the two transitions to the ES given here.

### 3. Methods of calculation

The total radiative capture cross sections σ(*NJ*,*J*$_f$) for the *EJ* and *MJ* transitions in the case of the PCM are given, for example, in works (Angulo et al. 1999; Dubovichenko 1995, 2012a, 2012b; Dubovichenko & Dzhazairov-Kakhramanov 1997) and they have the following form:

$$\sigma_c(NJ,J_f) = \frac{8\pi Ke^2}{\hbar^2 q^3} \frac{\mu}{(2S_1+1)(2S_2+1)} \frac{J+1}{J[(2J+1)!!]^2}$$
$$\times A_J^2(NJ,K) \sum_{L_i,J_i} P_J^2(NJ,J_f,J_i) I_J^2(J_f,J_i) \quad (2)$$

where σ – total radiative capture cross section; μ – reduced mass of initial channel particles; $q$ – wave number in initial channel; $S_1$, $S_2$ – spins of particles in initial channel; $K$, $J$ – wave number and momentum of γ-quantum in final channel; $N$ – is the $E$ or $M$ transitions of the $J$ multipole ordered from the initial $J_i$ to the final $J_f$ nucleus state.

The value $P_J$ for electric orbital $EJ(L)$ transitions has the form (see, e.g., Dubovichenko 2012a, 2012b; Angulo et al. 1999):

$$P_J^2(EJ,J_f,J_i) = \delta_{S_iS_f}\left[(2J+1)(2L_i+1)(2J_i+1)(2J_f+1)\right](L_i 0 J 0 | L_f 0)^2 \begin{Bmatrix} L_i & S & J_i \\ J_f & J & L_f \end{Bmatrix}^2,$$

$$A_J(EJ,K) = K^J \mu^J \left(\frac{Z_1}{m_1^J} + (-1)^J \frac{Z_2}{m_2^J}\right), \quad I_J(J_f,J_i) = \langle \chi_f | R^J | \chi_i \rangle. \quad (3)$$

Here, $S_i$, $S_f$, $L_f$, $L_i$, $J_f$, and $J_i$ – total spins, angular and total moments in initial (*i*) and final (*f*) channels; $m_1$, $m_2$, $Z_1$, $Z_2$ – masses and charges of the particles in initial channel; $I_J$ – integral over wave functions of initial $\chi_i$ and final $\chi_f$ states, as functions of cluster relative motion of n and $^{10}$B particles with intercluster distance $R$.

For consideration of the $M1(S)$ magnetic transition, caused by the spin part of magnetic operator (see, e.g., Ajzenberg & Grajner 1973), it is possible to obtain an expression (see, e.g., Dubovichenko 2012a, 2012b) using the following (see, e.g., Varshalovich, Moskalev & Khersonskii 1989):

$$P_1^2(M1,J_f,J_i) = \delta_{S_iS_f}\delta_{L_iL_f}\left[S(S+1)(2S+1)(2J_i+1)(2J_f+1)\right]\begin{Bmatrix} S & L & J_i \\ J_f & 1 & S \end{Bmatrix}^2,$$

$$A_1(M1,K) = \frac{e\hbar}{m_0 c} K\sqrt{3}\left[\mu_1 \frac{m_2}{m} - \mu_2 \frac{m_1}{m}\right], \quad I_J(J_f,J_i) = \langle \chi_f | R^{J-1} | \chi_i \rangle, \quad J=1. \quad (4)$$

Here, $m$ is the mass of the nucleus, and $\mu_1$ and $\mu_2$ are the magnetic moments of the clusters, the values of which are taken from Fundamental Physical Constants (2010) and Avotina & Zolotavin (1979).

The construction methods used here for intercluster partial potentials at the given orbital moment $L$, are expanded in (Dubovichenko 2012a, 2012b, 2013a) and here we will not discuss them further. The next values of particle masses are used in the given calculations: $m_n$ = 1.00866491597 amu (see, e.g., Fundamental Physical Constants 2010) and $m(^{10}B)$ = 10.012936 amu (see, e.g., Nuclear Wallet Cards database, 2010), and constant $\hbar^2/m_0$ is equal to 41.4686 MeV fm$^2$.

## 4. Interaction potentials

For all partial waves of the n$^{10}$B interaction potentials, i.e., for each partial wave with the given $L$, we used the Gaussian potential of the form:

$$V(r) = -V_0 \exp(-\alpha r^2). \quad (5)$$

Here, as mentioned before, we will not consider the influence of the first resonance at 0.17 MeV in the $^6S_{5/2}$ wave; therefore, we will use the potential with FSs leading to the zero scattering phase

$$V_0 = 160.5 \text{ MeV}, \alpha = 0.5 \text{ fm}^{-2}. \quad (6)$$

The $^6S_{5/2}$ scattering phase shift of this potential at energy up to 1.0 MeV is less than 0.5°. The same parameters we be used for the $^8S_{7/2}$ scattering wave, also ignoring the resonance.

The following parameters were obtained for the third resonance state $^6P_{5/2} + ^8P_{5/2}$ at 0.475(17) MeV:

$$V_0 = 106.615 \text{ MeV}, \alpha = 0.4 \text{ fm}^{-2}. \quad (7)$$

Such potential leads to resonance, i.e., the scattering phase shift equals 90.0°(1), at 475(1) keV (l.s.) with a width of 193(1) keV (c.m.), which is in good agreement with the data of reviews (Ajzenberg-Selove 1990; Kelley et al. 2012).

For the potential of the pure-by-spin GS of $^{11}$B in the n$^{10}$B channel, where the $^6P_{3/2}$ wave is used, the following parameters were obtained:

$$V_0 = 165.3387295 \text{ MeV}, \alpha = 0.45 \text{ fm}^{-2}. \quad (8)$$

We have obtained the value of the dimensionless AC = 1.53(1) in the range of 3–10 fm, the charged radius of 2.44 fm, and the mass radius of 2.39 fm at the binding energy of -11.454100 MeV with the accuracy of the finite-difference method, used for the calculation of the binding energy at $\varepsilon = 10^{-6}$ MeV (see, e.g., Dubovichenko 2012c). The AC error is connected with its averaging over the above-mentioned range of distances. The phase shift for such potential decreases smoothly until a value of 179° when the changes from zero to 1.0 MeV. The generalized Levinson theorem (see, e.g., Nemets et al. 1988) is used for the determination of the value of scattering phase shift at zero energy.

The AC value equal to 1.72 fm$^{-1/2}$ was obtained in Dolinskii, Mukhamedzhanov & Yarmukhamedov (1978) for the GS of $^{11}$B in the cluster channel n$^{10}$B, where the coefficient of neutron identity was assigned (see expression 83b in Blokhintsev, Borbei & Dolinskii 1977). In this work, a slightly different definition of AC was used, notably $\chi_L(r) = C \cdot W_{-\eta L+1/2}(2k_0 r)$ (with the Coulomb parameter $\eta$ equals zero, in this case), which is different from our previous works (Dubovichenko 2012a, 2012b) to the value $\sqrt{2k_0}$ that equals 1.19 fm$^{-1/2}$ for the GS; therefore, the AC value equals 1.44 in the dimensionless form. The improved value of 1.82(15) fm$^{-1/2}$ is given in the latest results for this AC (see, e.g., Yarmukhamedov 2013), and after re-computation, it gives 1.52(12) in the dimensionless form and agrees absolutely with the value for the GS potential of Eq. (8) obtained here.

The parameters of the GS potential and any BSs in the considered channel at the given number of the bound, allowed or forbidden states in the partial wave, are fixed quite unambiguously by the binding energy, the charge radius, and the asymptotic constant. The accuracy of the determination of the BS potential parameters is connected with the accuracy of the AC, which is usually equal to 10% to 20%. There are no another ambiguities in this potential, because the classification of the states according to the Young schemes allows us unambiguously to fix the number of BSs in this partial wave, which defines its depth completely, and the width of the potential depends wholly on the values of the charge radius and the AC.

The next parameters were obtained for the parameters of the $^6P_{5/2} + ^8P_{5/2}$ potential without FSs for the second ES of $^{11}$B in the n$^{10}$B channel with $J^\pi = 5/2^-$:

$$V_0 = 151.61181 \text{ MeV}, \alpha = 0.45 \text{ fm}^{-2}. \tag{9}$$

This potential leads to the binding energy of -7.0092 MeV at $\varepsilon = 10^{-4}$, which is completely coincident with the experimental value (see, e.g., Ajzenberg-Selove 1990; Kelley et al. 2012) for the charge radius of 2.44 fm, and the AC of 1.15(1) at the range of 3–13 fm.

The next parameters were obtained for the potential without FSs for the third ES $^6P_{3/2}$ pure-by-spin with $J^\pi = 3/2^-$:

$$V_0 = 149.70125 \text{ MeV}, \alpha = 0.55 \text{ fm}^{-2}. \tag{10}$$

These parameters lead to the binding energy of -6.4338 MeV at $\varepsilon = 10^{-4}$ coinciding with the experimental value (see, e.g., Ajzenberg-Selove 1990; Kelley et al. 2012); the AC equals 1.10(1) at the range of 3–13 fm, and the charge and mass radii are equal to 2.44 fm and 2.41 fm, respectively. The scattering phase shift for this potential decreases until 178° at the energy of 1.0 MeV.

The next parameters were obtained for the $^6P_{7/2} + {}^8P_{7/2}$ potential without FSs for the fourth ES of $^{11}$B in the n$^{10}$B channel with $J^\pi = 7/2^-$:

$$V_0 = 143.72353 \text{ MeV}, \alpha = 0.45 \text{ fm}^{-2}. \tag{11}$$

The binding energy of -4.7112 MeV at $\varepsilon = 10^{-4}$, which absolutely coincides with the experimental value (see, e.g., Ajzenberg-Selove 1990; Kelley et al. 2012) for the charge radius of 2.44 fm, and dimensionless AC equal to 0.94(1) at the range of 3–15 fm, was obtained with this potential. The scattering phase shift for this potential decreases smoothly until 178° at the energy of 1.0 MeV.

These parameters were obtained for the $^6P_{5/2} + {}^8P_{5/2}$ potential without FSs for the ninth ES of $^{11}$B in the n$^{10}$B channel with $J^\pi = 5/2^-$:

$$V_0 = 135.39620 \text{ MeV}, \alpha = 0.45 \text{ fm}^{-2}. \tag{12}$$

This potential leads to the binding energy of -2.5339 MeV at $\varepsilon = 10^{-4}$, which completely coincides with the experimental value (see, e.g., Ajzenberg-Selove 1990; Kelley et al. 2012) for the charge radius of 2.44 fm, and dimensionless AC equal to 0.70(1) at the range of 3–24 fm. The scattering phase shift for this potential decreases smoothly until 178° at the energy of 1.0 MeV.

### 5. The total cross section of the radiative neutron capture on $^{10}$B

The next experimental data were used for the comparison of the calculation results given in Figs. 1 and 2. The black points (●) show the total summed capture cross section from Igashira et al. (1994) at 23, 40, and 61 keV. The triangle (▲) represents the cross section of 500(200) μb from Bartholomew & Campion (1957) at the energy of 25 meV, and the open reverse triangle (∇) shows the new results for the cross section of 305(16) μb at 25 meV from Mughabghab (2006), given in review Kelley et al. (2012). It should be noted that other data for 390(11) μb, obtained in Firestone et al. (2008) and also shown in Figs. 1 and 2 by the open reverse triangle (∇), were published later – reference to these results is also given in review Kelley et al. (2012). The experimental measurements of work Igashira et al. (1994) for the transitions to different ESs of $^{11}$B are shown in Figs. 1 and 2: open circles (o) represent the total capture cross section to the GS $^6P_{3/2}$, open squares (□) represent the total capture cross section to the second ES $^{6+8}P_{5/2}$, black squares (■) represent the total capture cross section to the fourth ES $^{6+8}P_{7/2}$, and open triangles ($\Delta$) represent the total capture cross section to the ninth ES $^{6+8}P_{5/2}$. Furthermore, in Figs. 3–4, only part of these experimental results is given.

The $E1$ transition $^6S_{5/2} \to {}^6P_{3/2}$ from the $S$ scattering wave with potential of Eq. (6) to the GS with potential (8) was considered initially in our calculations, and the obtained capture cross section is shown in Fig. 1, represented by the short dashed line. The general dashed line shows the capture cross section to the second ES for potential of Eq. (9), identified in section 2 as No. 2. The dotted line shown at the bottom of Fig. 1 denotes the cross section of the transition $^6S_{5/2} \to {}^6P_{3/2}$ to the third ES of Eq. (10). The dot-dashed line shows the cross section of the transition to the fourth ES of Eq. (11), identified in section 2 as No. 4. The dot-dot-dashed line, which is almost superimposed with the dotted line, shows the transition from the $S$ scattering waves to the ninth ES with potential of Eq. (12). The solid line gives the total summed cross section of all the above considered transitions, which largely describes the experimental data for the total summed cross sections from works (Igashira et al. 1994; Bartholomew & Campion 1957) at the energy range from 25 meV to 61 keV correctly.

Let us note that in the measurements Igashira et al. (1994) the transition to the third ES is not taken into account, and as seen in Fig. 1, this leads to nearly the same cross section of the transition to the third ES and the ninth ES – dotted and dot-dot-dashed lines. Therefore, probably, it is necessary to add the cross section of the transition to the ninth ES to the total cross sections from Igashira et al. (1994) to obtain summed cross sections that are more correct, and this will be equivalent to taking into account the transition to the third ES. Such cross sections are shown in Figs. 1–4 by the open rhombus – this account influences weakly the total cross sections, which also agree with the results of our calculations.

As can be seen from the obtained results, the calculated line for the transition to the fourth ES is in a good agreement with the given black squares (experimental data) Igashira et al. (1994). The good agreement of the calculation, shown by the dot-dot-dashed line, can also be observed for the transition to the ninth ES, the experiment for which is shown by the open triangles (see, e.g., Igashira et al. 1994). The measurements for the transition to the second ES, shown in Fig. 1 by the open squares (see, e.g., Igashira et al. 1994), lie appreciably higher than the corresponding calculated line, shown by the dashed line. The measurements of the cross section for the transition to the GS, shown by the open circles (see, e.g., Igashira et al. 1994), lie much lower than the calculated line, shown by the short dashed line. Thereby, only two calculations conform to the experimental results for the transitions to the fourth and ninth ESs (see, e.g., Igashira et al. 1994), although the total summed cross sections, shown by the black points or rhombus, are described completely by the calculated line – the solid line in Fig. 1.

Because, we do not know the AC value for the second ES, it is always possible to construct the potential correctly describing the capture cross sections to this state, shown in Figs. 1 and 2 by the open squares (see, e.g., Igashira et al. 1994). For example, it is possible to use the potential with the parameters:

$$V_0 = 108.37443 \text{ MeV}, \alpha = 0.3 \text{ fm}^{-2}, \tag{13}$$

which leads to the binding energy of -7.0092 MeV, the charged radius of 2.44 fm, and the value of the AC equal to 1.45(1) at the range of 4–13 fm. The calculation results of the capture cross sections to this state from the $S$ scattering waves are shown in Fig. 2 by the dashed line, which is in a quite agreement with the experimental data (see, e.g., Igashira et al. 1994) shown by the open squares.

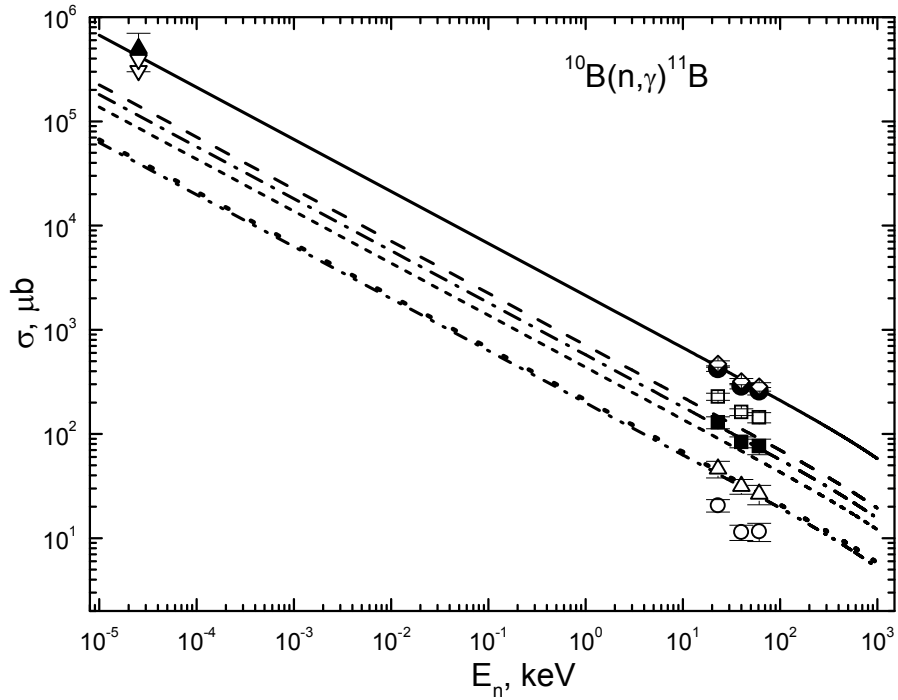

Fig. 1.– The total cross sections of the radiative neutron capture on $^{10}$B. Experimental data: black triangle (▲) – the capture cross section at 25 meV from Bartholomew & Campion (1957), points (●) – the total summed cross section of the neutron capture on $^{10}$B from Igashira et al. (1994), circles (o) – the total capture cross section to the GS, open squares (□) – the total capture cross section to the second ES, black squares (■) – the total capture cross section to the fourth ES, and open triangles (Δ) – the total capture cross section to the ninth ES from Igashira et al. (1994), open reversed triangles (∇) – the capture cross section at 25 meV from (Mughabghab 2006; Firestone et al. 2008), open rhombus (◊) – the summed total capture cross section from Igashira et al. (1994) taking into account the transition to the third ES. Lines: the short dashed line is the cross section of the $E1$ transition $^6S_{5/2} \to {}^6P_{3/2}$ from the $S$ scattering wave with potential of Eq. (6) to the GS with potential (4), the general dashed line is the capture cross section to the second ES of Eq. (9), the dotted line is the capture cross section of the transition $^6S_{5/2} \to {}^6P_{3/2}$ to the third ES of Eq. (10), the dot-dashed line is the cross section of the transition to the fourth ES of Eq. (11), the dot-dot-dashed line is the cross section of the transition from the $S$ scattering waves to the ninth ES with potential of Eq. (12), the solid line is the total summed cross section of all considered transitions.

At the same time, the other variant of the GS potential that describes the total capture cross sections to the GS correctly, shown in Figs. 1 and 2 by the open circles, will not agree with the known AC or that given above for the GS. For example, the parameters

$$V_0 = 602.548373 \text{ MeV}, \; \alpha = 2.0 \text{ fm}^{-2} \qquad (14)$$

allow one to describe reasonably the available experimental cross section measurements of the transition (see, e.g., Igashira et al. 1994), as is shown in Fig. 2 by the short dashed line. However, although this potential leads to the correct binding energy of -11.454100 MeV and describes reasonably the charged radius of 2.43 fm, the value of the AC is equal to 0.71(1) at the range of 2–8 fm, which is half that of the results from other experimental data (see, e.g., Dolinskii, Mukhamedzhanov & Yarmukhamedov 1978; Yarmukhamedov 2013). This result can be explained by the imperfection of the MPCM used here; however, on such occasions, the MPCM led to the correct description of the cross sections both to the transitions to the GS and to the total summed cross section of the capture processes (see, e.g., Dubovichenko 2012a, 2012b, 2013a; Dubovichenko, Dzhazairov-Kakhramanov & Burkova 2013). Therefore, it could be supposed that the experimental measurements for transitions to different ESs of $^{11}$B at the radiative neutron capture on $^{10}$B should be improved in the future; it will also be interesting to obtain new data in the range of possible resonances from 100 to 600 keV.

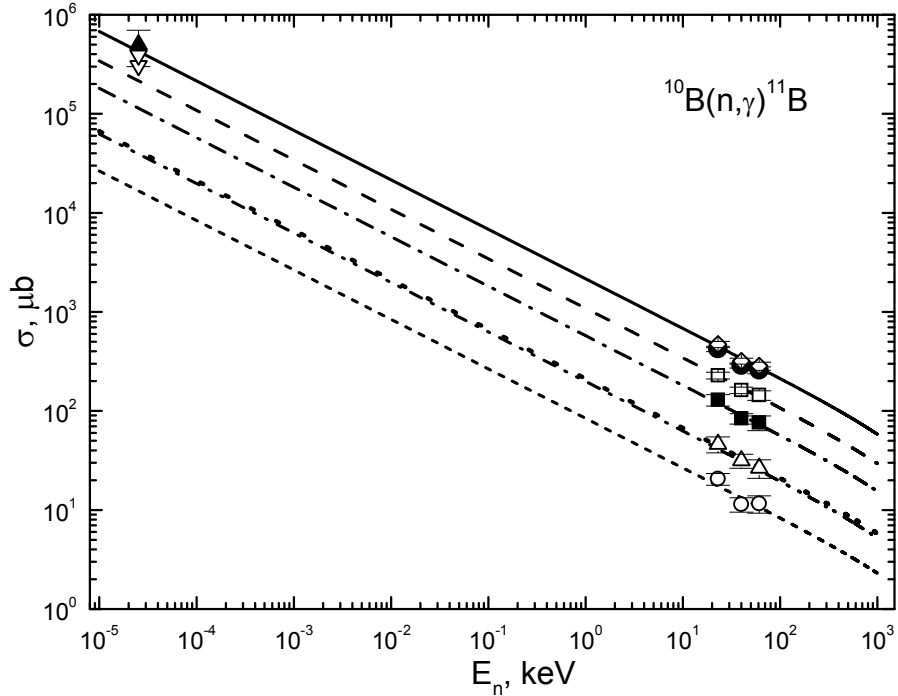

Fig. 2.– The total cross sections of the radiative neutron capture on $^{10}$B. Experimental data: the same as in Fig. 1. Lines: the short dashed line is the cross section of the $E1$ transition $^6S_{5/2} \to {}^6P_{3/2}$ from the $S$ scattering wave with potential of Eq. (6) to the GS with potential of Eq. (14), the general dashed line is the capture cross section to the second ES of Eq. (13), the dotted line is the capture cross section of the transition $^6S_{5/2} \to {}^6P_{3/2}$ to the third ES of Eq. (10), the dot-dashed line is the cross section of the transition to the fourth ES of Eq. (11), the dot-dot-dashed line is the cross section of the transition from the $S$ scattering waves to the ninth ES with potential of Eq. (12), the solid line is the total summed cross section of all considered transitions.

Reverting to the calculation results given in Fig. 1, we note that at the energies from 10 meV to 10 keV, the calculated cross section is almost a straight line, and it can be approximated by a simple function of the form:

$$\sigma_{ap} = \frac{A}{\sqrt{E_n}}. \qquad (15)$$

The value of the given constant $A = 2123.4694$ μb·keV$^{1/2}$ was determined from a single point of the cross-sections (solid line in Fig. 1) at a minimal energy of 10 meV. The absolute value $M(E) = \left|[\sigma_{ap}(E) - \sigma_{theor}(E)]/\sigma_{theor}(E)\right|$ of the relative deviation of the calculated theoretical cross-sections ($\sigma_{theor}$), and the approximation of this cross-section ($\sigma_{ap}$) by the expression given above in the energy range until 10 keV, is at the level of 0.2%. It is supposed that this form of total cross-section dependence on energy will be conserved at lower energies. In this case, the estimation of the cross-section value, for example, at the energy of 1 μ keV, gives the value of 67.2 b. The coefficient for the solid line in Fig. 2 in the expression given above in Eq. (15) for the approximated calculation results for cross sections, is equal to 2150.3488 μb·keV$^{1/2}$.

Furthermore, the considered $M$1 transitions to the GS and to the different ESs are shown in Fig. 3, together with the summed cross section for the $E$1 processes, which is shown by the dashed line (it is represented by the solid line in Fig. 1). The dotted line at the top of the figure shows the cross section of the $M$1 transition to the GS with potential of Eq. (8) from the resonant $^6P_{5/2}$ scattering wave for potential of Eq. (7), identified in section 2 as No. 6. The dot-dashed line it is the transition from the $P_{5/2}$ scattering wave of Eq. (7) to the second ES with potential of Eq. (9), identified in section 2 as No. 7. The dot-dot-dashed line shows the cross section of the $M$1 transition $^6P_{5/2} \rightarrow {}^6P_{3/2}$ to the third ES with potential of Eq. (10) in Fig. 3.

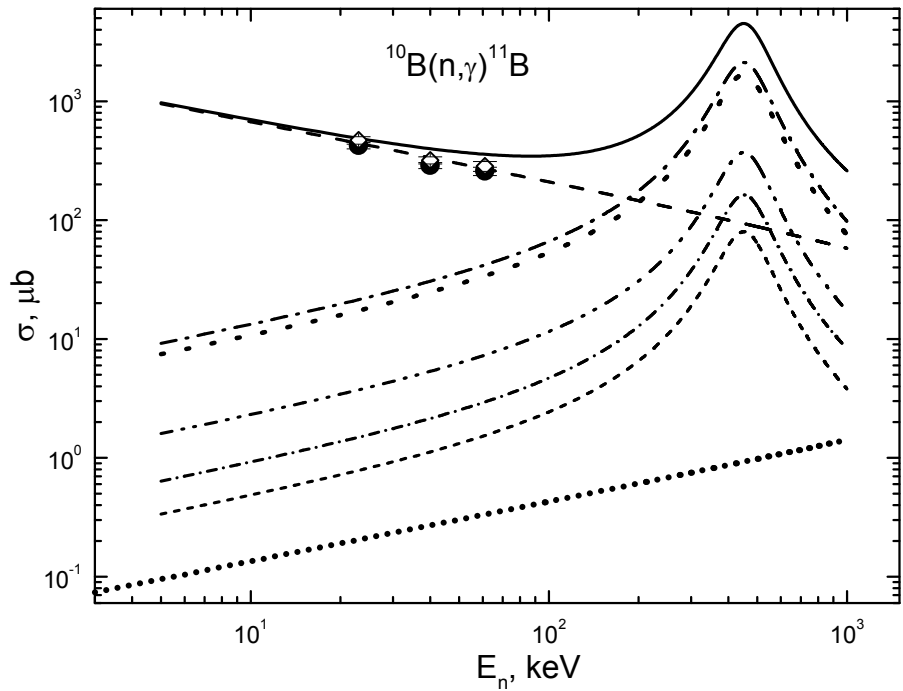

Fig. 3.– The total cross sections of the radiative neutron capture on $^{10}$B. Experimental data: points (●) – the total summed cross section of the neutron capture on $^{10}$B from Igashira et al. (1994), open rhombus (◊) – the summed total capture cross section from Igashira et al. (1994) taking into account the transition to the third ES. Lines: the dashed line is the summed cross section of the $E$1 transitions, shown in Fig. 1 by the solid line, the dotted line gives the cross section of the $M$1 transition to the GS of Eq. (8) from the resonant $^6P_{5/2}$ scattering wave for potential of Eq. (7), the dot-dashed line is the cross section of the transition from the resonant $^6P_{5/2}$ scattering wave to the second ES of Eq. (9), the dot-dot-dashed line shows the cross section of the $M$1 transition $^6P_{5/2} \rightarrow {}^6P_{3/2}$ to the third ES of Eq. (10), from the $S$ scattering waves to the ninth ES with potential of Eq. (12), the short dashed line is the $M$1 transition to the fourth ES of Eq. (12), the dot-dashed line with closely placed dashes shows the ninth ES of Eq. (8), the dotted line with closely placed dots at the bottom of the figure shows the $M$1 transition from the non-resonant $^6P_{3/2}$ scattering wave to the GS, the solid line shows the sum of the $E$1 and $M$1 transitions considered above.

Another possible $M$1 transition to the fourth ES of Eq. (11), identified in section 2 as No. 9, has the form of the cross section shown in Fig. 3 by the short dashes. In addition, the *M*1 transition to the ninth ES of Eq. (12) is possible; it is identified in section 2 as No. 10, and shown in Fig. 3 by the dot-dashed line with closely placed dashes (the third line at the bottom), which lies slightly higher than the short dashed line. In addition, the *M*1 transition to the GS from the non-resonant $^6P_{3/2}$ scattering wave was considered with the potential of zero depth – the second transition under No. 6 in section 2. The result is shown by the dotted line with closely placed dots in the bottom of Fig. 3. Its value at the maximum is about 1.5 μb and it has practically no influence on the calculated cross sections in the range of the resonance at 475 keV, almost reaching to 4.5 mb.

The sum of all the *E*1 and *M*1 transitions described above is shown in Fig. 3 by the solid line, which gives a suitable description of the given experimental data. The small overshoots of the calculated cross sections over the experimental one at 40 and 61 keV can be used to argue that the used potential of Eq. (7) leads to the overestimated value of the resonance width in the $P_{5/2}$ scattering wave of 193 keV. As mentioned before, some values for the energy and width of this resonance are given in review Ajzenberg-Selove (1990), and it will be possible to construct new potentials, which will be matched with the width 31 keV, as shown in Table 11.11 from Ajzenberg-Selove (1990). The use of such potentials can change the results for the resonance cross sections, reducing their influence to the total summed calculated cross sections in the energy range 40–60 keV. We will use the resonance potential of the $P_{5/2}$ scattering wave in the form

$$V_0 = 3555.983 \text{ MeV}, \alpha = 13.0 \text{ fm}^{-2}. \tag{16}$$

This potential, as before, leads to the resonance at 475 keV, but its width is reduced until 32 keV (c.m.), in full accordance with the data listed in Table 11.11 of work Ajzenberg-Selove (1990).

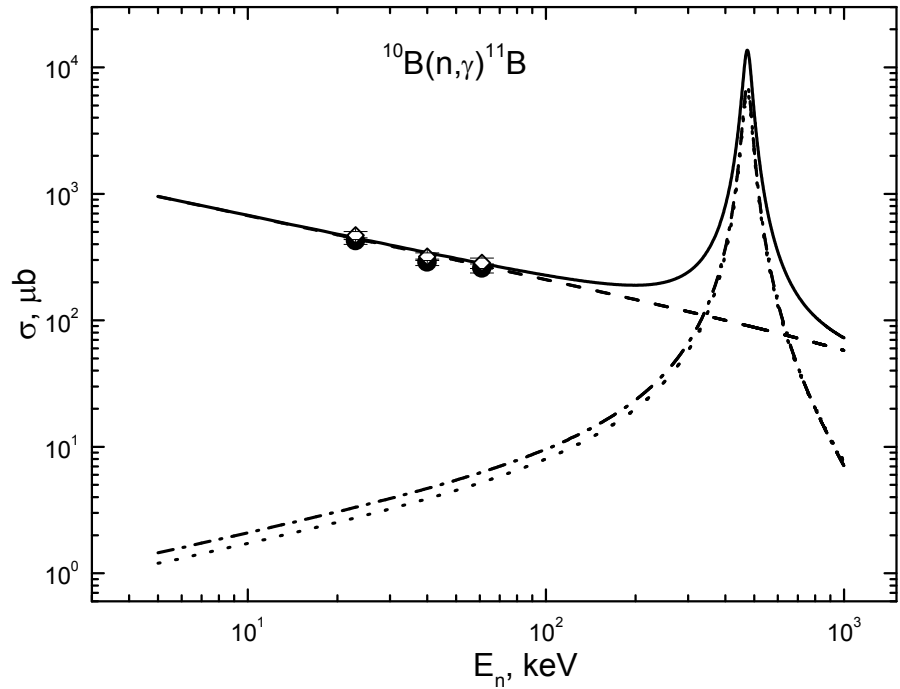

Fig. 4.– The total cross sections of the radiative neutron capture on $^{10}$B. Experimental data: the same as in Fig. 3. Lines: the dashed line is the capture cross section for the *E*1 processes shown in Fig. 1 by the solid line, the dotted line shows the cross section of the *M*1 transition to the GS of Eq. (8) from the resonance $^6P_{5/2}$ scattering wave for potential of Eq. (16), the dot-dashed line is the cross section of the transition from the resonance $P_{5/2}$ scattering wave of Eq. (16) to the second ES with potential of Eq. (9), the solid line shows the total summed cross section.

The calculation results of the total cross sections with potential of Eq. (16) are shown in Fig. 4. The dashed line is the summed cross section for the $E$1 processes, shown in Fig. 1 by the solid line. The cross section of the $M$1 transition to the GS with potential of Eq. (8) from the resonance $^6P_{5/2}$ scattering wave for potential of Eq. (16) is shown by the dotted line, and the dot-dashed line shows the cross section from the resonance $P_{5/2}$ scattering wave of Eq. (16) to the second ES with potential of Eq. (9). The solid line is the summed cross sections for all considered transitions. All other transitions shown in Fig. 3 lead to the cross sections that do not provide an essential contribution to the total summed cross sections at the resonance energy. As can be seen from Fig. 4, the resonance part of the calculated cross section does not really change the total summed cross sections at 61 keV, which now are in a good agreement with the available experimental data of Igashira et al. (1994).

## 6. Conclusion

As can be seen from the listed results, the obvious assumptions about the methods of construction of the n$^{10}$B interaction potentials, if they have FSs, allow one to obtain acceptable results on the description of the available experimental data for the total cross section of the neutron capture on $^{10}$B (see, e.g., Igashira et al. 1994; Bartholomew & Campion 1957; Firestone et al. 2008) at the energy range from 25 meV to 61 keV. The possibility to describe all considered experimental data both by capture cross section and according to the GS characteristics, allows us to fix parameters of the GS potential closely enough in the form of Eq. (8). The summed cross sections at the resonance energy of 0.475 MeV, equals 4.5 μb at the width of the resonance of 193 keV and 13.7 μb at the width of 32 keV.

Thereby, the MPCM again confirms, as already done in 25 reactions (see, e.g., Dubovichenko 2012a, 2012b, 2013a, 2013b, 2013c, 2013d, 2014a, 2014b; Dubovichenko et al. 2014; Dubovichenko, Dzhazairov-Kakhramanov & Burkova 2013; Dubovichenko & Dzhazairov-Kakhramanov 2009, 2012a, 2012b; Dubovichenko, Dzhazairov-Kakhramanov & Afanasyeva 2013; Dubovichenko & Uzikov 2011), its ability to describe correctly the cross sections of the processes such as the radiative capture of neutral and charged particles on light nuclei at thermal and astrophysical energies. As this occurs, such results are obtained using the potentials matched with the resonance scattering phases, or with the level spectra of the final nucleus and the BS characteristics of the considered nuclei, and some basic principles of the construction of such potentials were checked partially in the three-body calculations (see, e.g., Dubovichenko, S.B. 2011).

## Acknowledgments

The work was performed under grant No. 0151/GF2 "Studying of the thermonuclear processes in the primordial nucleosynthesis of the Universe" of the Ministry of Education and Science of the Republic of Kazakhstan.
In conclusion, the authors express their deep gratitude to Prof. Yarmukhamedov R. (INP, Tashkent, Uzbekistan) for provision of the information on the AC in the n$^{10}$B channel, and also to Prof. Strakovsky I.I. (GWU, Washington, USA) and Blokhintsev L.D. (Moscow State University, Moscow, Russia) for extremely useful discussions on certain parts of this work.